\documentclass[conference]{IEEEtran}
\IEEEoverridecommandlockouts
\usepackage{cite}
\usepackage{amsthm}

\usepackage{hyperref}

\usepackage{verbatim}
\usepackage{booktabs}

\usepackage{amsmath,amssymb,amsfonts}
\usepackage{algorithmic}
\usepackage[ruled,vlined,algo2e,linesnumbered]{algorithm2e}
\usepackage{setspace}
\usepackage{subfigure}
\usepackage{pifont}
\usepackage{graphicx}
\usepackage{textcomp}
\usepackage{xcolor}
\def\BibTeX{{\rm B\kern-.05em{\sc i\kern-.025em b}\kern-.08em
    T\kern-.1667em\lower.7ex\hbox{E}\kern-.125emX}}
\begin{document}

\title{Scalable Quantum Neural Networks for Classification\\
}

\author{\IEEEauthorblockN{Jindi Wu}
\IEEEauthorblockA{\textit{Department of Computer Science} \\
\textit{William \& Mary}\\
Williamsburg, VA, USA \\
jwu21@wm.edu}
\and
\IEEEauthorblockN{Zeyi Tao}
\IEEEauthorblockA{\textit{Department of Computer Science} \\
\textit{William \& Mary}\\
Williamsburg, VA, USA \\
ztao@cs.wm.edu}
\and
\IEEEauthorblockN{Qun Li}
\IEEEauthorblockA{\textit{Department of Computer Science} \\
\textit{William \& Mary}\\
Williamsburg, VA, USA \\
liqun@cs.wm.edu}

}

\maketitle

\begin{abstract}
Many recent machine learning tasks resort to quantum computing to improve classification accuracy and training efficiency by taking advantage of quantum mechanics, known as quantum machine learning (QML). The variational quantum circuit (VQC) is frequently utilized to build a quantum neural network (QNN), which is a counterpart to the conventional neural network. Due to hardware limitations, however, current quantum devices only allow one to use few qubits to represent data and perform simple quantum computations. The limited quantum resource on a single quantum device degrades the data usage and limits the scale of the quantum circuits, preventing quantum advantage to some extent. To alleviate this constraint, we propose an approach to implementing a scalable quantum neural network (SQNN) by utilizing the quantum resource of multiple small-size quantum devices cooperatively. In an SQNN system, several quantum devices are used as \textit{quantum feature extractors}, extracting local features from an input instance in parallel, and a quantum device works as a \textit{quantum predictor}, performing prediction over the local features collected through classical communication channels. The quantum feature extractors in the SQNN system are independent of each other, so one can flexibly use quantum devices of varying sizes, with larger quantum devices extracting more local features. Especially, the SQNN can be performed on a single quantum device in a modular fashion. Our work is exploratory and carried out on a quantum system simulator using the TensorFlow Quantum library. The evaluation conducts a binary classification on the MNIST dataset. It shows that the SQNN model achieves a comparable classification accuracy to a regular QNN model of the same scale. Furthermore, it demonstrates that the SQNN model with more quantum resources can significantly improve classification accuracy.

\end{abstract}

\begin{IEEEkeywords}
Quantum Machine Learning, Quantum Neural Networks, Variational Quantum Circuits, Distributed Quantum computing
\end{IEEEkeywords}

\section{Introduction}\label{sec:introduction}

Quantum machine learning (QML) is a revolutionary approach combining machine learning (ML) and quantum computing \cite{schuld2015introduction, biamonte2017quantum, houssein2022machine}. The explosion of data volume and memory consumption makes the ML algorithms run on classical computers unsupportable, although the massive data is desired for ML algorithms to train models. On the contrary, quantum computing based on the law of quantum mechanics (e.g., superposition, entanglement and teleportation) excels at efficiently processing information and outperforms classical computing \cite{nielsen2002quantum, steane1998quantum}. For example, Shor's algorithm is sub-exponentially faster in factoring \cite{shor1999polynomial} and Grover's algorithm is quadratically faster in searching \cite{grover1996fast}. Therefore, the combination of ML and quantum computing is a natural trend. 

\begin{figure}[t]
\centering 
\includegraphics[width=.9\columnwidth]{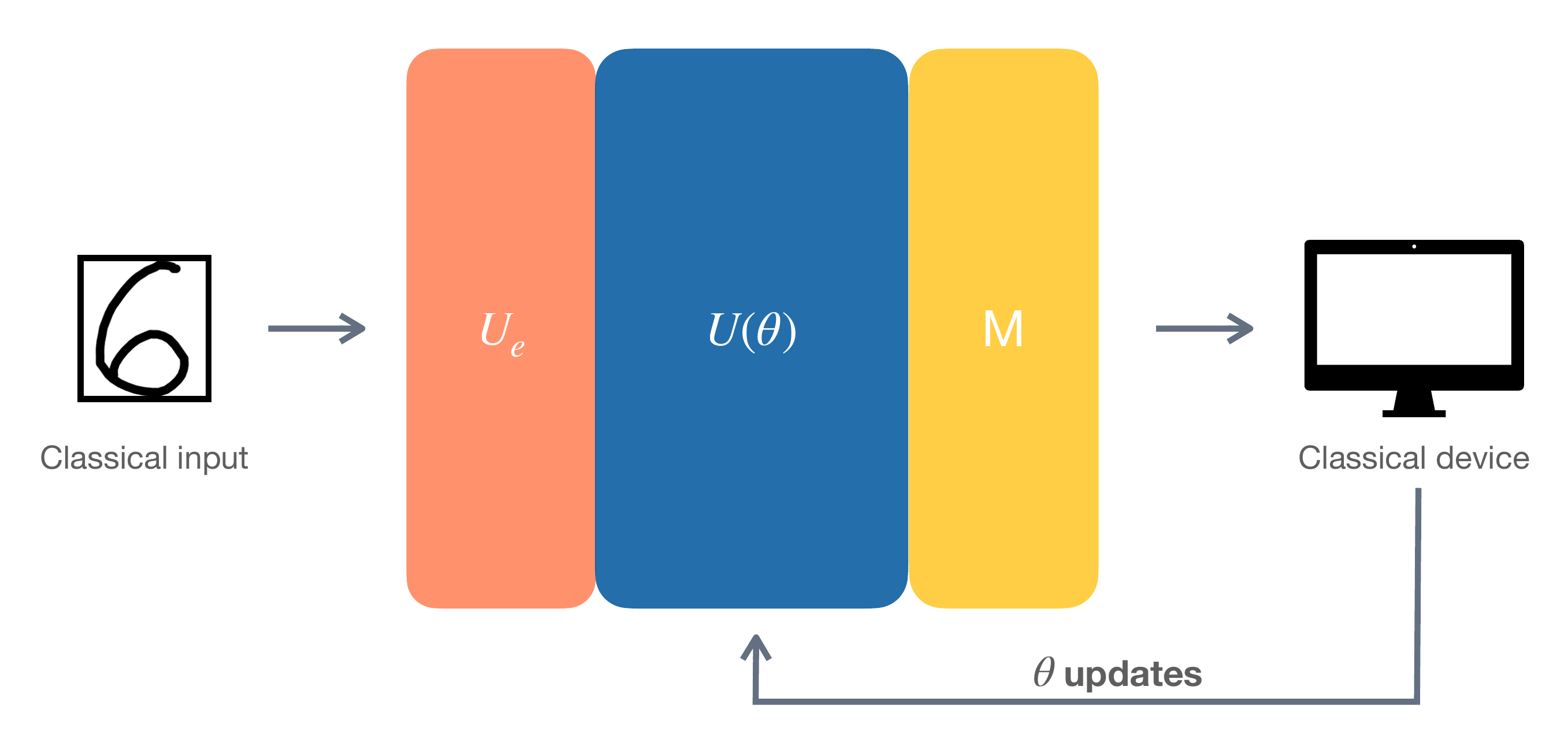}
\caption{General QNN in hybrid quantum-classical architecture. The encoding unit $U_e$ prepares quantum states for classical input data. The VQC $U(\theta)$ process the prepared quantum states with variational parameters $\theta$, which will be updated by a classical device according to the circuit output measured by $M$.}
\label{fig:generalqnn}
\end{figure}

Numerous applications of classical neural networks (NNs) have achieved huge success, which motivates many research into quantum neural networks (QNNs) \cite{beer2020training, killoran2019continuous, farhi2018classification, zhou2007quantum, zhou2010quantum, adhikary2020supervised}. Considering the current Noisy Intermediate-Scale Quantum (NISQ) devices are prone to many noises, the variational quantum circuit (VQC) is the general approach to building QNNs on such devices \cite{cerezo2021variational, preskill2018quantum}. VQC, which consists of a set of quantum gates with trainable parameters, is a counterpart of classical NN made up of neurons. The QML algorithm implemented with VQC is the hybrid quantum-classical method that jointly employs quantum and classical devices, as shown in Fig.~\ref{fig:generalqnn}. This method trains a QNN model on a classical dataset by running its quantum circuit on a quantum device, which prepares qubits and modifies their quantum states using quantum gates (unitary transformations). And based on the measurement results of the qubits, calculating the updates to VQC's parameters on a classical device using a classical optimizer, e.g., stochastic gradient descent (SGD) \cite{bottou2012stochastic, sweke2020stochastic}. More details of QNN will be provided in Sec.~\ref{sec:preliminary}.


A barrier to achieving quantum supremacy is the limited number of qubits and connectivities on NISQ devices. With such limitations, only small-scale QNNs can be constructed, and they are unable to load the high-dimensional classical data. Hence, many efforts have been made to overcome this limitation, including quantum encoding and data dimension reduction methods. The amplitude encoding method can load $2^n$-dimensional classical data on $n$ qubits by taking the advantage of quantum entanglement and superposition, but its hardware implementation is inefficient and may induce additional errors. Angle encoding is an alternative method that has easier hardware implementation, but it can only load $n$-dimensional data on $n$ qubits. The encoding methods involve a trade-off between representation capability and hardware implementation. Therefore, quantum encoding is still an open problem. Moreover, Chen et al. use a pre-trained classical NN to extract features from original data in order to fit the data dimension to the number of available qubits on quantum devices \cite{chen2021federated}. And Steni et al. use principal component analysis (PCA) \cite{lloyd2014quantum} to downscale original data \cite{stein2022quclassi}. Nevertheless, These methods that require additional computation undermine the advantages of quantum acceleration. This unsolved problem motivates us to design a QNN system that is not limited by the size of current quantum devices and is easy to scale up for high-dimensional classical data. We name it as Scalable Quantum Neural Network (SQNN).

An SQNN system combines the quantum resources of multiple small-size quantum devices to simulate a large quantum device with sufficient resources and trains a large-scale QNN for high-dimensional data. Suppose there are five available small-size quantum devices in a quantum system for an image (high-dimensional data) classification task. Four of them work as \textit{quantum feature extractors} and the remaining one acts as a \textit{quantum predictor}. The SQNN system partitions a training instance into four segments that can fit the capacity of the quantum feature extractors and assigns them to quantum feature extractors. Each quantum feature extractor encodes the received classical data to quantum data, uses a VQC to reduce the original data size and represent the information as abstract features, then obtains the extracted features by measuring the readout qubits of the circuit. The quantum predictor collects extracted local quantum features from quantum feature extractors via classical communication channels, learns from them, and makes prediction on them with a VQC. So far, the SQNN system could obtain the prediction result by measuring the readout qubits of the quantum predictor's VQC. Then like the optimization of the classical NNs, a classical device calculates the updates of the parameters in VQCs according to the prediction results and the pre-defined objective function. The updates are further used to set the VQCs' parameters for the next training step. 

The SQNN system enables the QNN models to learn from high-dimensional data by scaling QNNs up. Furthermore, the SQNN system supports a flexible data partition strategy to fully utilize the quantum system's resources. With the strategies, SQNN assigns the high-dimensional data segments of different sizes to quantum feature extractors according to their various amount of quantum resources. Fig.~\ref{fig:mnist6} illustrates the possible input partition strategies. Especially, a quantum system with fewer or even a single quantum device can also train an SQNN. After partitioning the training instance into several segments, one can repeatedly use the available quantum devices as quantum feature extractors to act on data segments with recording the intermediate result, then as the quantum predictor.

The major contributions of this paper as following:

\begin{itemize}
    \item We present the first, to the best of our knowledge, scalable quantum neural network (SQNN) that can learn from high-dimensional data without being constrained by the quantum hardware limitation on qubit amount.
    \item We propose a scalable quantum ML approach that collaboratively uses the quantum resources of multiple small-size quantum devices.
    \item We implement a SQNN for a binary classification task and conduct extensive evaluations to show the effectiveness of the proposed design.
\end{itemize}

In the rest of this paper, we will recap some background knowledge about quantum gates, quantum encoding methods, and QNN in Sec.~\ref{sec:preliminary}. We will introduce the proposed approach in Sec.~\ref{sec:sqnn}, and show the evaluations in Sec.~\ref{sec:exp}. Some related works will be reviewed in Sec.~\ref{sec:related}. At last, we will conclude our work in Sec.~\ref{sec:concl}.

\section{Preliminary}\label{sec:preliminary}

\subsection{Quantum gates}
Qubit is the basic unit of quantum computation. Compared with a classical bit that can only represent state 0 or 1, a qubit can simultaneously represent state 0 and 1 with a certain probability distribution. The state of a qubit is denoted as
\begin{equation*}
    |\psi\rangle = \alpha |0\rangle + \beta |1\rangle \text{ with } \alpha, \beta \in \mathbb{C}
\end{equation*}
where $|\alpha| ^ 2$ and $|\beta|^2$ respectively express the probability of the qubit is measured as state 0 and 1, i.e., $|\alpha| ^ 2 + |\beta|^2 = 1$. The qubit state is also written as a vector $|\psi\rangle = [\alpha, \beta]^\top$.

The quantum computation is conducted by changing the state of the qubits with quantum gates $U$ (unitary matrices that satisfy $U^\dagger U = U U^\dagger = I$). The Pauli gates \{$X$, $Y$, $Z$\} and the Hadamard gate $H$ is the fundamental single-qubit gate. Their matrices are shown as
\begin{equation*}
    \begin{split}
    X&= \left[ \begin{array}{ccc}
    0 & \ \ 1\\
    1 & \ \ 0\end{array} \right]\qquad
    Y= \left[ \begin{array}{ccc}
    0 & -i\\
    i & 0\end{array} \right]\\
    Z &= \left[ \begin{array}{ccc}
    1 & 0\\
    0 & -1\end{array} \right]\quad
    H= \frac{1}{\sqrt{2}}\left[ \begin{array}{ccc}
    1 & 1\\
    1 & -1\end{array} \right]
    \end{split}
\end{equation*}
The Pauli gates $X$, $Y$ and $Z$ rotate the qubit by $\pi$ radians around the X, Y and Z-axis of the Bloch sphere (a visual representation of quantum state). And the Hadamard gate $H$ creates an equal superposition state for the computational basis states: $|0\rangle \to (|0\rangle + |1\rangle) / \sqrt{2}$ and $|1\rangle \to (|0\rangle - |1\rangle) / \sqrt{2}$. It is a $\pi / 2$ rotation around the Y-axis followed by a $\pi$ rotation around the X-axis.

Based on the Pauli gates, the single-qubit rotation gates are generated that allow the qubit to be rotated arbitrary radians in the Bloch sphere. The $R_x(\theta)$, for example, rotates the qubit by $\theta$ radians around the X-axis in the Bloch sphere as
\begin{equation*}
    R_x(\theta) = \exp\left(-i X \frac{\theta}{2}\right) = \left[ \begin{array}{ccc}
    \cos(\frac{\theta}{2}) & -i \sin(\frac{\theta}{2})\\
    -i \sin(\frac{\theta}{2}) & \cos(\frac{\theta}{2})\end{array} \right]
\end{equation*}
The same for $R_y(\theta)$ and $R_z(\theta)$. Ising coupling gates are two-qubit gates expended from single-qubit rotation gates, e.g.,
\begin{equation*}
\begin{split}
    R_{xx}(\theta) &= \exp\left(-i X \otimes X \frac{\theta}{2}\right) \\
    &= \left[ \begin{array}{cccc}
    \cos(\frac{\theta}{2}) & 0 & 0 & -i \sin(\frac{\theta}{2})\\
    0 &  \cos(\frac{\theta}{2}) & -i \sin(\frac{\theta}{2}) & 0 \\
    0 & -i \sin(\frac{\theta}{2}) & \cos(\frac{\theta}{2}) & 0\\
    -i \sin(\frac{\theta}{2}) &  0 & 0 & \cos(\frac{\theta}{2})\end{array} \right]
\end{split}
\end{equation*}
It performs the same rotation on two qubits simultaneously. And the same for $R_{yy}(\theta)$ and $R_{zz}(\theta)$. In QML, Ising coupling gates are frequently used to construct VQC \cite{broughton2020tensorflow}.

\subsection{Quantum encoding} \label{sec:back_encoding}
To be processed on quantum devices, classical data must be encoded into quantum states in Hilbert space. Because current quantum devices have limitations on the number of qubits and the depth of circuits, the encoding methods impact the efficiency of hardware implementation and the design of the following quantum information processing circuits. This subsection will introduce three mainly used quantum encoding schemes \cite{schuld2018supervised}. 

\subsubsection{Basis encoding}
The basis encoding method is the most straightforward quantum encoding method for arithmetic operation. Basis encoding first represents classical data in binary form and directly maps them onto quantum computational bases. For example, a numerical data point [0.3, 0.6, 0.2, 0.8] will be converted into a binary data point [0, 1, 0, 1] based on a threshold 0.5, and mapped to the 4-qubit quantum computational basis $|0101 \rangle$. For the implementation of the encoding circuit, four qubits are prepared with initialized state $|0\rangle$, and an Pauli-$X$ gate is appended after the second and fourth qubits to flip their states to $|1\rangle$. 
In general, the basis encoding method takes $n$ qubits to load the binary representation of a classical data point $x = (x_1, x_2, \cdots, x_n)$ and encodes it as 
\begin{equation*}
    |\Psi_x\rangle = \bigotimes_i^n |x_i\rangle
\end{equation*}
where $\bigotimes$ is tensor product operator.

\subsubsection{Angle encoding}
The angle encoding method loads the classical data as the radians of rotation gates acting on qubits. $n$ qubits and $n$ quantum rotation gates $R \in \{R_x, R_y, R_z\}$ are required to embed the classical data in size $n$. For a classical data instance $x = (x_1, x_2, \cdots, x_n)$, the angle encoding method prepares it as
\begin{equation*}
    |\Psi_x\rangle = \bigotimes_i^n R(x_i) |0\rangle
\end{equation*}

\subsubsection{Amplitude encoding}
The amplitude encoding method embeds classical data into the amplitudes of a quantum state. An $n$-dimensional normalized classical vector $x = (x_1, x_2, \cdots, x_n)$ is encoded to the amplitudes of $\log n$-qubit quantum state by
\begin{equation*}
    |\Psi_x\rangle = \sum^{n}_{i=1}x_i |i\rangle
\end{equation*}
where $|i\rangle$ is the $i$-th computational basis state and $\sum|x_i|^2 = 1$.

\begin{figure*}[h]
\centering 
\includegraphics[width=.8\linewidth]{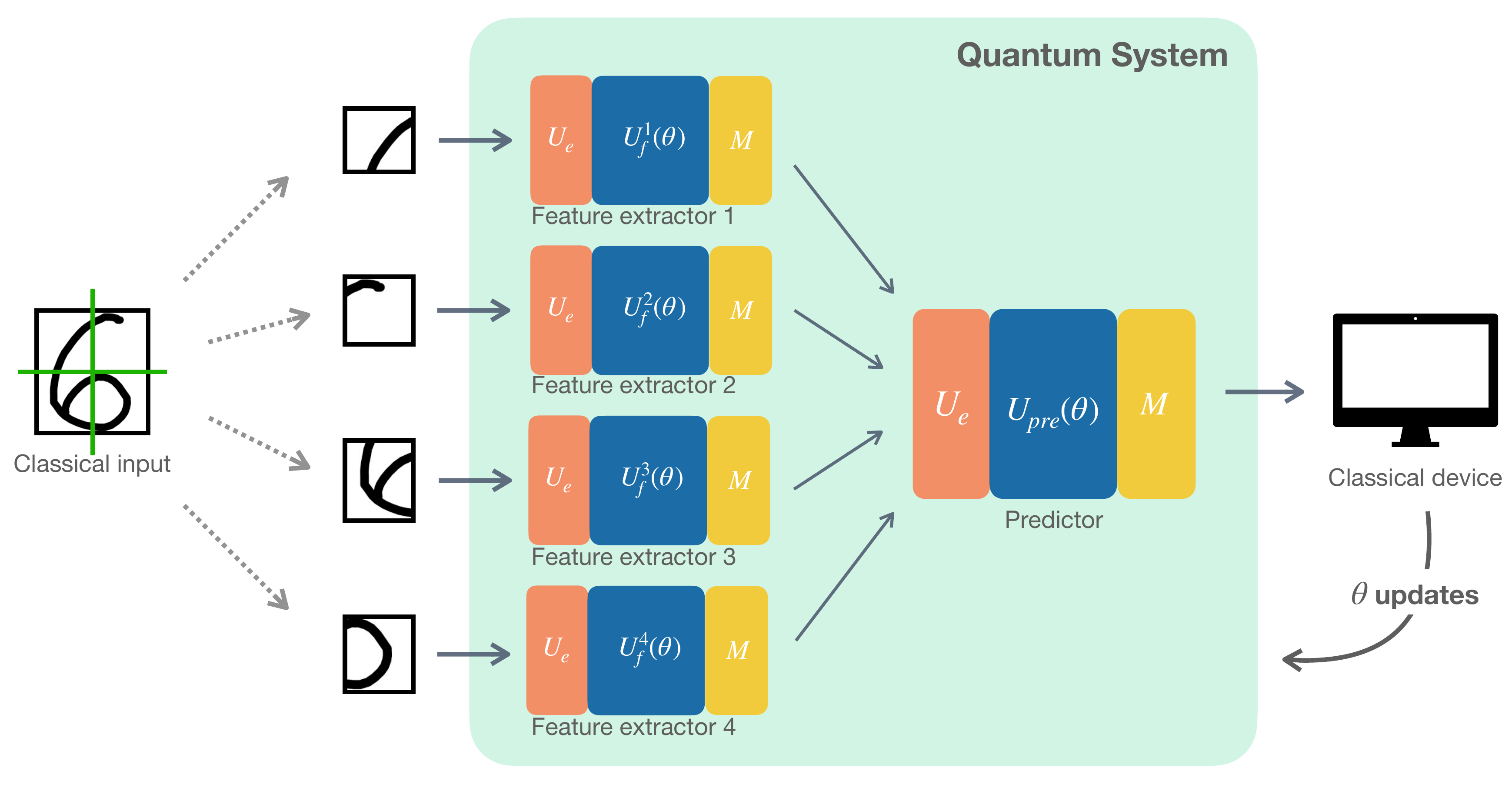}
\caption{An SQNN in hybrid quantum-classical architecture. The quantum system contains five small-size quantum devices, four of which work as quantum feature extractors to extract features from the segments of input data, and the remaining one as a predictor to make predictions using the extracted features collected from feature extractors. Each quantum device has a quantum circuit that consists of an encoding unit $U_e$, a VQC $U(\theta)$, and a measurement unit $M$. The classical device will assist them in updating their variational parameters during training.  }
\label{fig:sqnn}
\end{figure*}

\subsection{Quantum neural network}
The current widely used QNN that can run on NISQ devices for classification tasks was proposed in 2018 \cite{farhi2018classification}, which is an analogue of the classical NN in the quantum computing field. The QNN is a quantum circuit within a sequence of parameter-dependent quantum gates (unitary operators) which act on quantum input data. Generally, a QNN can be shown as 
\begin{equation}
    U(\theta) = \prod_{l=1}^N V_l U_l(\theta_l)
\end{equation}
which is a product of $N$ quantum layers \cite{broughton2020tensorflow}. The $l$-th quantum layer consists of product of non-parametric quantum gates $V_l$ and parametric quantum gates $U_l(\theta_l)$ where $\theta_l$ are variational parameters. We can further represent the parametric quantum gates $U_l(\theta_l)$ in $l$-th layer as the production of $S$ parametric quantum gates,
\begin{equation}
    U_l(\theta_l) \equiv \bigotimes_{j=1}^S U_{l, j}(\theta_{l, j})
\end{equation}
In which each parametric quantum gate $U_{l, j}(\theta_{l, j})$ can be transformed with Euler's formula as 
\begin{equation}
    \exp(-i \theta_{l, j} P) = I \cos(\theta_{l, j}) - i \sin(\theta_{l, j})P
\end{equation}
where $i$ is the imaginary number, $I$ is a 2 $\times$ 2 identity matrix, and $P$ is a Pauli operator from the set \{$X$, $Y$, $Z$\} that acts on qubits.

The output of the QNN is the measurement result on a computational basis of the readout qubits. Since the measurement result of a qubit is probabilistic, the expectation value $E$ of the measurement results is used as the QNN output,
\begin{equation} 
    \label{eq:measure}
    E = \langle \Psi_x| U^\dagger (\theta) M U (\theta) |\Psi_x\rangle
\end{equation}
where $ |\Psi_x\rangle$ is the input quantum state of the QNN and $M$ is a linear combination of Pauli operators that serve as observables for readout qubits.

The loss $L$ of a training sample in hybrid quantum-classical model is calculated in conventional manner on a classical device. For the given training sample, the loss $L$ is calculated with an objective function $\ell(.)$ of the task on the expected output $y$ and the actual output $E$
\begin{equation}
    L = \ell(E, y)
\end{equation}

During the model optimization phase, like in the classical NN, back-propagation and gradient descent will be performed to update variational parameters in the QNN. The gradient of a variational parameter $\theta_k$ in $k$-th quantum layer with respect to with respect to loss $L$ can be calculated by
\begin{equation}
    \frac{\partial L}{\partial \theta_k} = \frac{\partial L}{\partial E} \frac{\partial E}{\partial \theta_k}
\end{equation}
It is easy to obtain $\partial L / \partial E$ according to the objective function $\ell(.)$. $\partial E / \partial \theta_k$ could be calculated by
\begin{equation}
    \label{eq:grads}
    \frac{\partial E}{\partial \theta_k} = i \langle \Psi_x|U^\dagger_-[P_k, U^\dagger_+ H U_+] U_- | \Psi_x \rangle
\end{equation}
where 
\begin{equation}
    U_+ = \prod_{l=k+1}^N V_l U(\theta_l)
    \mbox{ and }
    U_- = \prod_{l=1}^{l=k} V_l U(\theta_l).
\end{equation}
With the gradients of parameters, the classical device sets updated parameters for QNN using an optimization algorithm, such as SGD. An alternative gradient calculation for a quantum model is parameter-shift, which obtains the gradients by running the same VQCs with shifted parameters and calculating the difference in their outputs \cite{mitarai2018quantum, schuld2019evaluating}.

\section{Scalable Quantum Neural Network}~\label{sec:sqnn}

The limited quantum resources on NISQ devices constrain the scalability of quantum circuits and computational power of quantum computing, particularly for QML tasks that demand extensive computation. In this section, we present a solution, SQNN, to circumvent the hardware constraints of quantum devices. The concept behind SQNN is to construct and train a large-scale VQC by cooperatively using the quantum resource on multiple small-size quantum devices. The architecture overview of the proposed design is shown in Fig.~\ref{fig:sqnn}. The components of SQNN are detailed in the following subsections. The adopted encoding method is discussed in \ref{sec:encode}. The variational quantum layers of SQNN are introduced in \ref{sec:layer}. The optimization of SQNN is described in \ref{sec:optmization}. \ref{sec:single} shows how to build SQNN in small quantum systems. And the important notations are listed in Table~\ref{tab-notation}.

Considering a quantum system with $p+1$ small-size quantum devices, we use $p$ of them as \textit{quantum feature extractors} for quantum local feature extraction, and the remaining one as a \textit{quantum predictor} for prediction (classification). Given a classical training dataset in which the instances $x$ are too large to be loaded by a quantum device in the system, we partition $x$ into $p$ small segments of size $n$ using the first partition strategy

\begin{figure}[t]
\centering 
\includegraphics[width=.9\columnwidth]{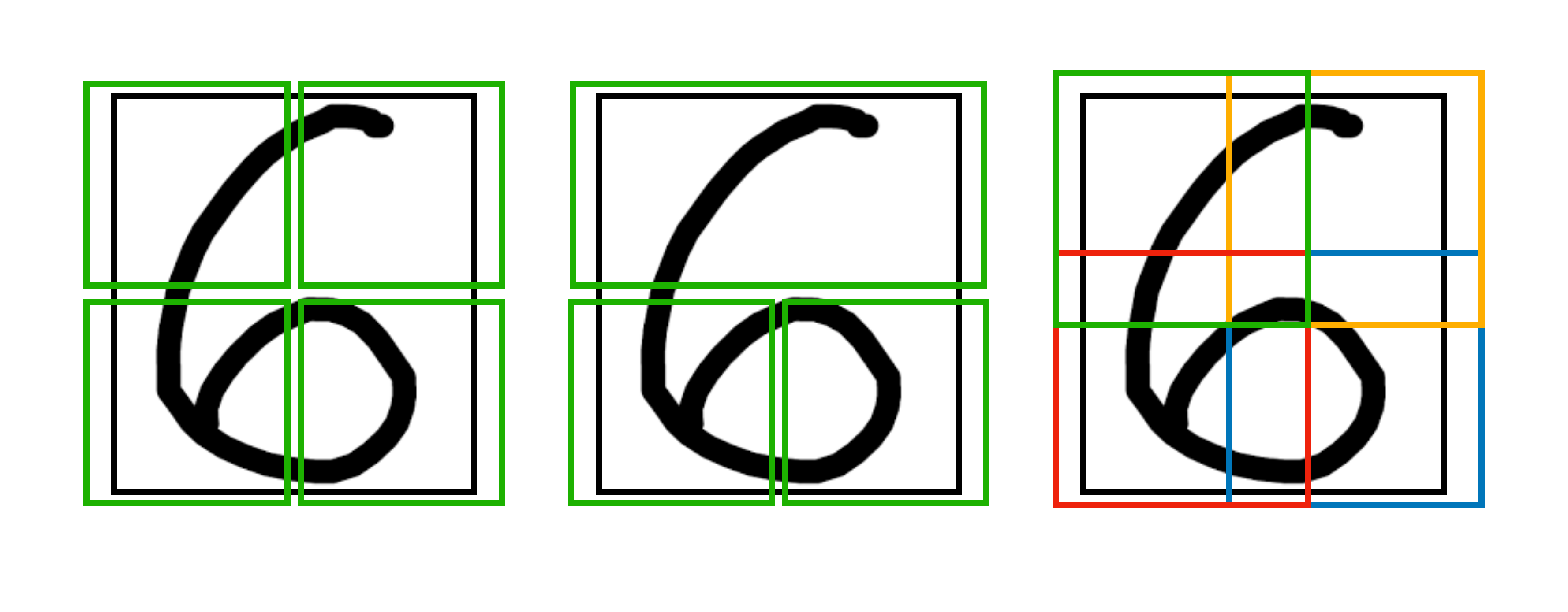}
\caption{Input partition strategies. Depending on the size of quantum feature extractors, a high-dimensional input image could be evenly partitioned without overlap (left), unevenly partitioned without overlap (middle), or evenly partitioned with overlap (right).}
\label{fig:mnist6}
\end{figure}

\begin{table}[h]
\caption{Notation list}
\label{tab-notation}
\begin{center}
\begin{tabular}{l p{6.5cm}}
\toprule 
\textbf{Notation}  & \textbf{Description} \\
\midrule
$x$  &  A classical training instance\\
$|\Psi_x\rangle$ & The quantum state representation of $x$\\
$f_i$  &  The $i$-th quantum feature of a training instance\\
$y$ & The correct label of the training instance\\
$y'$ & The predicted label of the training instance\\
$\theta$ & The variational parameters of VQC\\
$R(\theta)$ & A rotation gate with $\theta$ radians\\
$U_f^i(\theta)$  &  The VQC on $i$-th quantum feature extractor\\
$U_{pre}(\theta)$  &  The VQC on quantum predictor\\
$U_{e}(x)$  &  The quantum circuit of the encoding unit\\
$M$ & The qubit measurement unit\\
\bottomrule
\end{tabular}
\end{center}
\end{table}

\noindent shown in Fig.~\ref{fig:mnist6}, where $n$ is less than the number of available qubits on the quantum devices. We denote the instance in segments as $x = \{x^1, x^2,  \cdots, x^p\}$ and the $i$-th segment of instance $x$ as $x^i = \{x^i_1, x^i_2, \cdots, x^i_n \}$.

\subsection{Encoding unit}\label{sec:encode}

Both the quantum feature extractors and the quantum predictor in the SQNN system need the encoding unit to transform classical data into quantum data for further quantum operations. We use the angle encoding method in the encoding unit because of its simple hardware implementation and ability to support the gradient calculation for input quantum data, which is necessary for SQNN to perform back-propagation via chain rule.

There are some reasons that the other commonly used encoding methods are not adopted in our work. As mentioned in the Sec.~\ref{sec:back_encoding}, the basis encoding method embeds numerical data into binary data and is implemented by selectively appending Pauli-X gate behind the qubits with initial state $|0\rangle$. Although the basis encoding method is straightforward, it is not adopted for several reasons. First, it processes the original data with a discontinuous function that does not support the gradient calculation for the input of VQC. Second, the binary representation of an instance will lose much useful information, which is not desirable in ML tasks. And third, multiple numerical instances may map to the exact binary representation, which reduces the amount of training data. The amplitude encoding method is also not used in this work because its hardware implementation is complex and does not offer a simple way to calculate the gradients of quantum circuit input. The details of the calculation and usage of input gradients of quantum circuits in SQNN will be discussed in the Sec.~\ref{sec:optmization}.

\begin{figure}[t]
\centering 
\includegraphics[width=.72\linewidth]{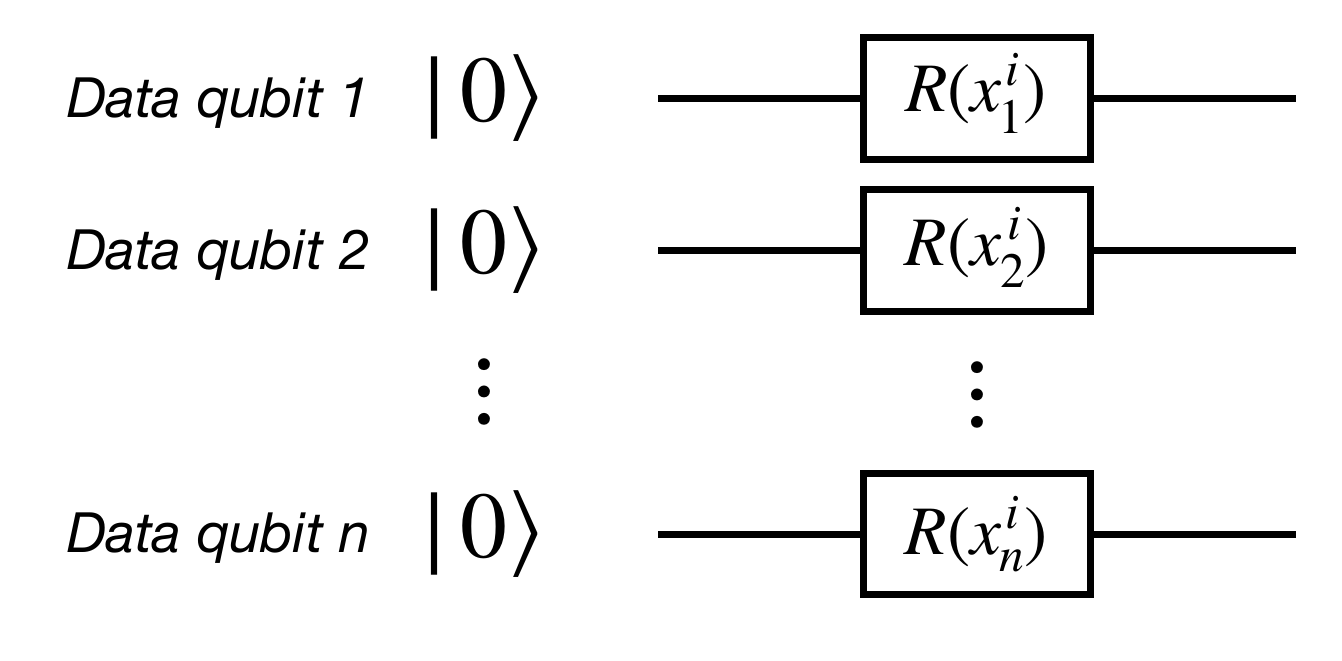}
\caption{Angle encoding unit. The classical data vector $x^i = [x^i_1, x^i_2, \cdots, x^i_n]$ is loaded on $n$ data qubits using rotation gate $R \in \{R_x, R_y, R_z\}$.}
\label{fig:encoding}
\end{figure}

The circuit of the quantum angle encoding unit is illustrated in Fig~\ref{fig:encoding}. When a quantum device receives the $i$-th segment of the classical instance $x$, it flats the segment as a vector $x^i = [x^i_1, x^i_2, \cdots, x^i_n ]$ and maps the entries into $[0, 2\pi)$, and prepares $n$ qubits with initial sate $|0\rangle$. The encoding unit sets the state of qubits independently to represent the data using rotation gate $R$. To be more specific, $j$-th qubit will pass a rotation gate $R(x_j^i)$ to load $x_j^i$ in its state as a certain superposition
\begin{equation}
    |\Psi_{x_j^i}\rangle =  R(x_j^i) |0\rangle = \cos\left(\frac{x^i_j}{2}\right) |0\rangle - i \sin\left(\frac{x^i_j}{2}\right)|1\rangle
\end{equation}
Hence, the prepared quantum state for $x^i$ is shown as a tensor product of each qubit state
\begin{equation}
    |\Psi_{x^i}\rangle = \bigotimes_j^n |\Psi_{x^i_j}\rangle.
\end{equation}

\subsection{Variational quantum layers} \label{sec:layer}

\begin{figure*}[!htb]
\centering
\subfigure[Quantum feature extraction layer]{\includegraphics[width=.9\columnwidth]{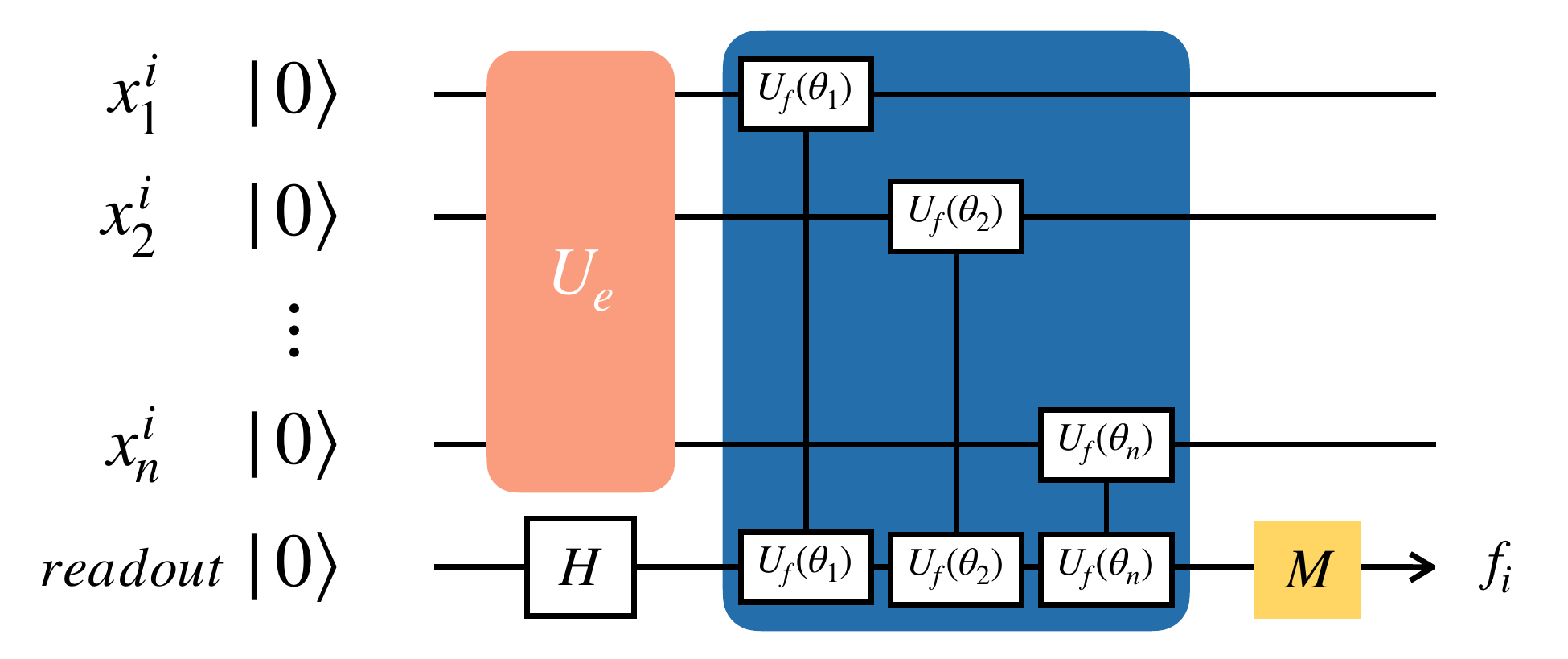}\label{fig:layers:extractor}}
\subfigure[Quantum prediction layer]{\includegraphics[width=.9\columnwidth]{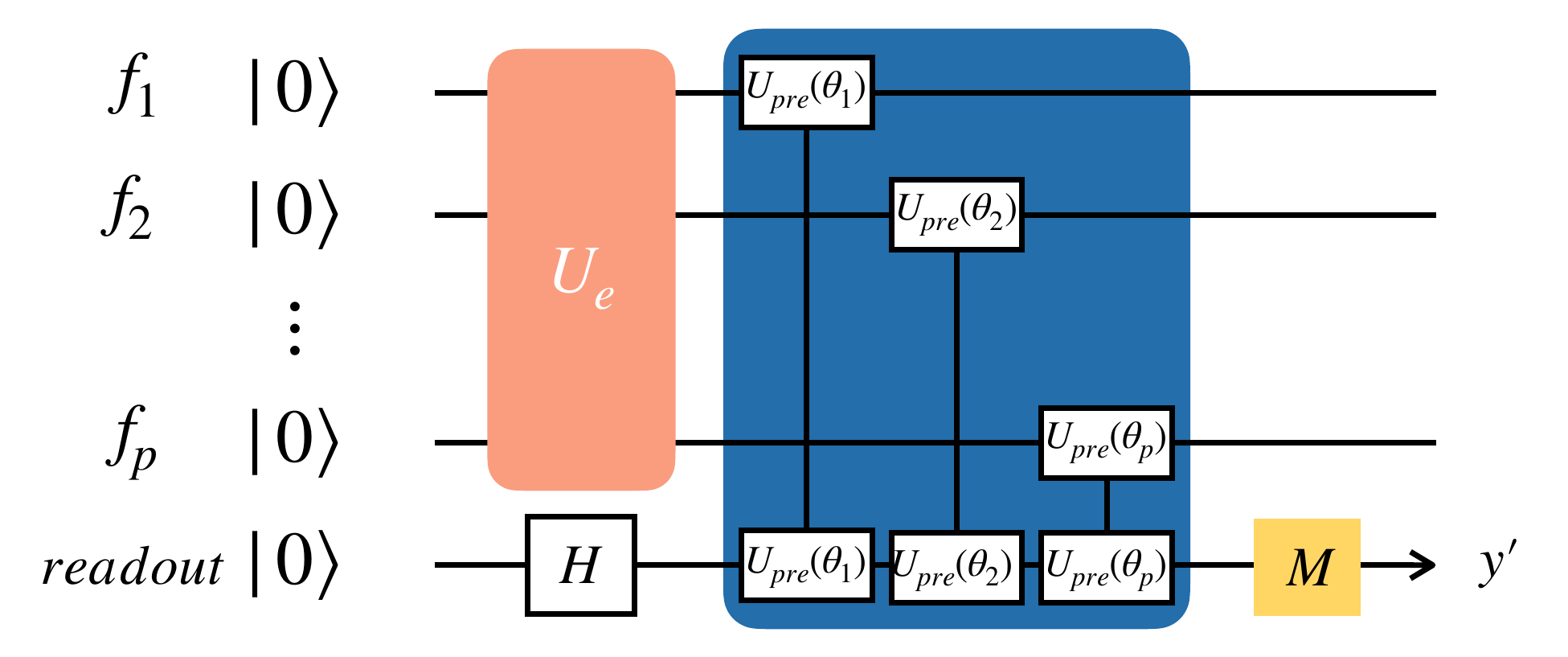}\label{fig:layers:predictor}}
\caption{Two types of quantum layers in SQNN}
\label{fig:layers}
\end{figure*}

A SQNN consists of two types of variational quantum layers: quantum feature extraction layer and quantum prediction layer. They are deployed on quantum feature extractors and a quantum predictor, respectively.

\subsubsection{Quantum feature extractor}

A quantum feature extractor that receives $i$-th piece of the classical instance $x^i$ of size $n$ will use $n+1$ qubits in its quantum circuits, as shown in Fig.~\ref{fig:layers:extractor}. The first $n$ qubits are \textit{data qubits} that pass the encoding unit to load the classical data as quantum state $|\Psi_{x^i}\rangle$. And the last one serves as \textit{readout qubit} that will be measured to obtain the results of the quantum data processing. In general, a quantum circuit could have multiple readout qubits, but we only show one out of simplification.

The quantum feature extraction layer is a VQC that maps the quantum features from the data qubits to the readout qubit using two-qubit parameterized quantum gates that create entanglement between them. In this work, we adopt the Ising coupling gates to entangle qubits and follow the same design of VQC introduced in \cite{tfqweb}. We define the $n$ gates that consecutively act on $n$ pairs of data and readout qubits as a \textbf{block}. Similar to the classical NN, the variational gates works as neurons and a block as a layer of the QNN. The block could be repeated to build a complex QNN with more parameters. We denote the quantum feature extraction layer deployed on $i$-th quantum feature extractor as $U_f^i(\theta_f^i)$ where $\theta_f^i$ is the trainable parameters. With the entanglement with all data qubits, the state of the readout is the extracted quantum features from the segment $x^i$ and will be obtained as a real number $f_i$ by measurement with Eq.~\ref{eq:measure}

\begin{equation}
    \label{eq:feature}
    f_i = \langle \Psi_{x^i}| {U_f^i}^ \dagger (\theta_f^i) M U_f^i(\theta_f^i) |\Psi_{x^i}\rangle
\end{equation}

\subsubsection{Quantum predictor}

The quantum predictor collects extracted local features that represented as real numbers $f = \{f_1, f_2, \cdots, f_p\}$ from $p$ quantum feature extractors via conventional communication channels. The circuit of the predictor uses $p$ data qubits to load extracted local features using the encoding unit and one readout qubit for outputting prediction result, as shown in Fig.~\ref{fig:layers:predictor}. Generally, the number of readout qubits depends on the classification task, i.e., $\log k$ readout qubits are needed for a $k$-class classification task. Assuming we are performing a binary classification task, only one readout qubit is used. 

The quantum prediction layer is a VQC that jointly learns local features and makes a prediction on them. The design of the VQC is the same as that of the quantum feature extraction layer. We represent the quantum prediction layer as $U_{pre}(\theta_{pre})$ and $\theta_{pre}$ stands for the trainable parameters. The prediction result of the instance $x$ is
\begin{equation}
    \label{eq:pred}
    y' = \langle \Psi_f| {U_{pre}}^ \dagger (\theta_{pre}) M U_{pre}(\theta_{pre})|\Psi_{f}\rangle
\end{equation}
where $|\Psi_f \rangle$ is the quantum state of the extracted features $f$ and $y'$ is a real number in the range of [-1, 1]. For the binary classification with labels -1 and 1, we map the result $y'$ to label -1 if it is negative; otherwise, to label 1.

\subsection{Optimization}~\label{sec:optmization}
After the forward propagation, a classical device receives the prediction result $y'$ of $x$ from the quantum predictor. The classical device will calculate the loss of the current training instance $x$ by
\begin{equation}
    L = \ell(y, y')
\end{equation}
where $\ell(\cdot)$ is a pre-defined loss function, e.g., Mean squared error (MSE) loss, and $y$ is the actual label of $x$. The classical device with the knowledge of VQCs on quantum devices will optimize the parameters of SQNN. In the following, we will demonstrate the gradient derivation of the quantum prediction layer and quantum feature extraction layers according to back-propagation and the optimization with the gradient descent method.

The gradient of the variational parameters $\theta_{pre}$ in the quantum prediction layer with respect to the loss $L$ is obtained by
\begin{equation}
    \frac{\partial L}{\partial \theta_{pre}} = \frac{\partial L}{\partial y'} \frac{\partial y'}{\partial \theta_{pre}}
\end{equation}
where the $\partial y'/ \partial \theta_{pre}$ can be calculated with Eq.~\ref{eq:grads} and $\partial L / \partial y'$ is easily calculated. Then the parameters $\theta_{pre}$ will be updated with the learning rate $r$ by
\begin{equation}
    \theta_{pre} = \theta_{pre} - r \frac{\partial L}{\partial \theta_{pre}}
\end{equation}

The gradients of the variational parameters $\theta_f^i$ in the $i$-th quantum feature extraction layer with respect to the loss $L$ can be obtained as follow
\begin{equation}
    \frac{\partial L}{\partial \theta_f^i} = \frac{\partial L}{\partial y'} \frac{\partial y'}{\partial f_i} \frac{\partial f_i}{\partial \theta_f^i}
\end{equation}
The $\partial f_i / \partial \theta_f^i$ can be calculated by Eq.~\ref{eq:grads} as well. The $\partial y' / \partial f_i$ is the partial derivative of the input local feature $f_i$ with respect to the output of the quantum predictor $y'$. With the encoding method mentioned above, the input $f_i$ of the quantum prediction layer is encoded in the $i$-th qubit by using $R(f_i)$. Hence, we have
\begin{equation}
    \frac{\partial y'}{\partial f_i} = i \langle 0|[R, {U_{pre}}^\dagger(\theta_{pre}) M {U_{pre}(\theta_{pre})}] | 0 \rangle.
\end{equation}
To optimize $i$-th quantum feature extraction layer, the classical device updates $\theta_f^i$ with learning rate $r$ by
\begin{equation}
    \theta_f^i = \theta_f^i - r \frac{\partial L}{\partial \theta_f^i}.
\end{equation}

So far, we have gone through the optimization process of SQNN on a single training instance. To train the SQNN in the mini-batch style, one just needs to record the gradients of parameters for each sample in a mini-batch of data, and make the classical device update the variational parameters using the average of gradients.

\subsection{SQNN in small quantum systems}\label{sec:single}

This section shows how the SQNN can be built and trained in a small quantum system with insufficient quantum devices with the assistance of a classical device. We consider a particular case where only one quantum device is available in the quantum system.

Since the quantum feature extractors work independently, we partition the classical training instance into $p$ segments of the equal size based on the number of qubits on the quantum device, and make the device serve as $p$ feature extractors sequentially. When the quantum device works as the $i$-th quantum feature extractor, it builds the circuit $U_f^i$ with initialized parameters $\theta^i_f$ and performs quantum operations on the data segment $x^i$. The classical device records the circuit structure, current parameters, and the extracted local feature $f_i$ for it. Once the device has completed the work as quantum feature extractors, it acts as the quantum predictor by initializing the quantum predictor circuit $U_{pre}(\theta_{pre})$ and performing it on the extracted features $f$. The classical device also records the circuit structure and parameters for it. The classical device then updates the parameters as introduced in \ref{sec:optmization}. In the following training steps, the quantum device will first reproduce the circuits and reload the latest parameters before repeating the previous step. 

In the quantum system that consists of several quantum devices of different sizes, one can divide the classical instance into segments that fit the various sizes of available quantum devices as Fig.~\ref{fig:mnist6}, to fully use the system's quantum resources.\\

\section{Experiments and results}\label{sec:exp}
The SQNN is evaluated using a binary classification task. We conduct the binary classification using the images labeled ``3" and ``6"  in the MNIST handwritten digits dataset \cite{deng2012mnist}. After removing the images with other labels, we have 12049 training and 1968 test samples. The handwritten digit images in MNIST are pre-processed to position the digits in the center of $28 \times 28$ fields. In our experiments, we simulate the circuits of SQNN on a classical device with the TensorFlow Quantum library. And two scenarios of multi-quantum machine systems for SQNN are considered: the SQNNs with quantum feature extractors of the same and different sizes.

Here we define the form of the \textbf{basic model} used in our experiments to construct QNN classifiers. For a quantum device with $n+1$ qubits, the basic model is defined as: the first $n$ qubits are data qubits that store quantum data, and the last qubit is the readout that provides the quantum computation result by measurement. In each \textbf{block} of the basic model, the entanglement between each data and the readout qubit is created by the same variational Ising coupling gates in $\{R_{xx}(\theta), R_{yy}(\theta), R_{zz}(\theta)\}$ as shown in Fig.~\ref{fig:layers}. The source code used to generate the experiment results is available in \href{https://github.com/Jindi0/SQNN.git}{github.com/Jindi0/SQNN}.

\subsection{Observations} \label{sec:obs}

\begin{figure*}[ht]
    \centering
    \subfigure[Training loss and validation accuracy of 3-block QNN classifiers]{
    \label{fig:small_3}
    \includegraphics[width=.9\linewidth]{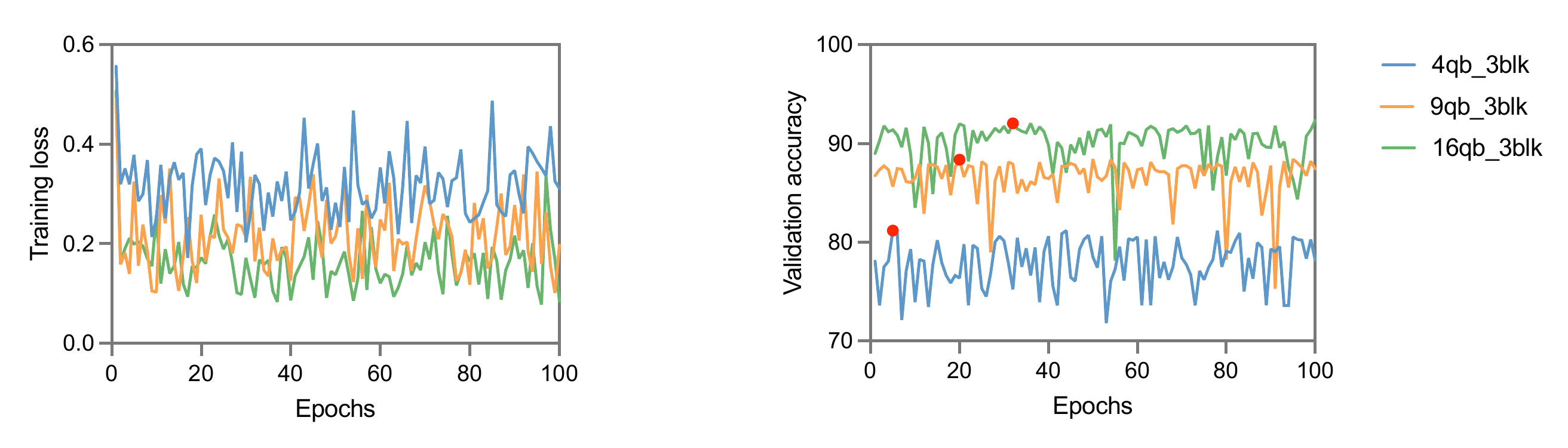}
    }\\
    \subfigure[Training loss and validation accuracy of 6-block QNN classifiers]{
    \label{fig:small_6}
    \includegraphics[width=.9\linewidth]{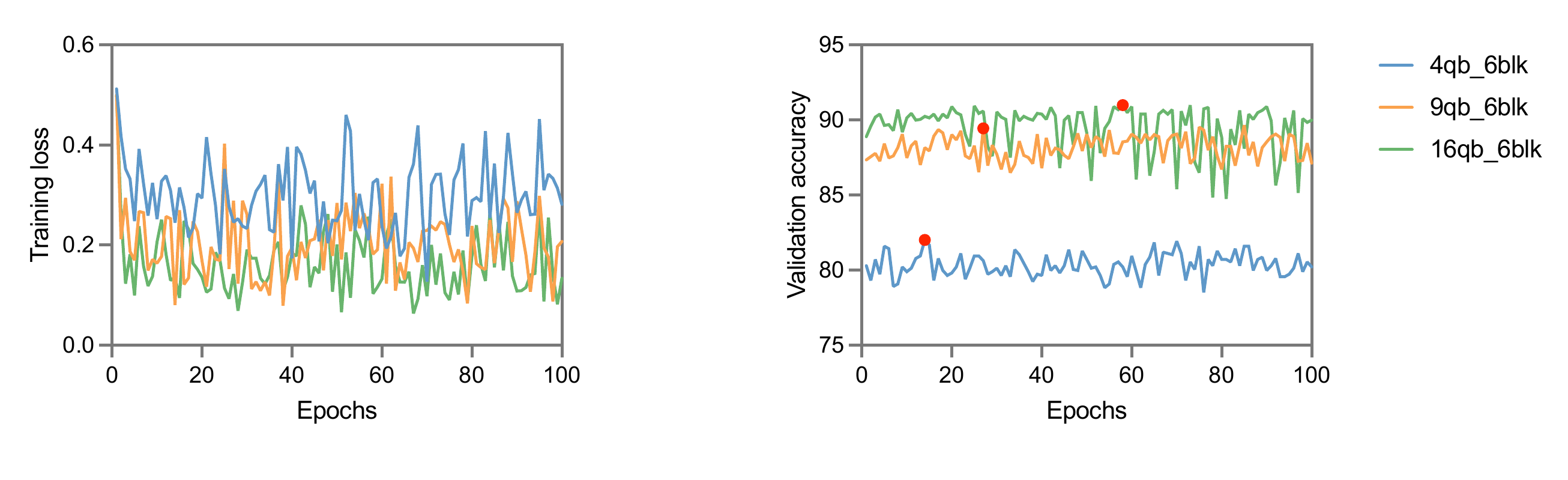}
    }
    \caption{\textbf{The performance of QNN classifiers on a single device.} The single-device QNN models consist of 4, 9, and 16 data qubits (qb). The QNNs in each size are evaluated using three and six variational circuit blocks (blk). Their best accuracy is marked with red dots during 100 epochs of training. These results indicate the QNN model that learns from higher dimensional data can achieve better performance.}
    \label{fig:small_qubits}
\end{figure*}

\begin{figure*}[!h]
    \centering
    \subfigure[Accuracy of 4-qubit QNNs classifiers]{
    \label{fig:small_comp_4qb}
    \includegraphics[width=.6\columnwidth]{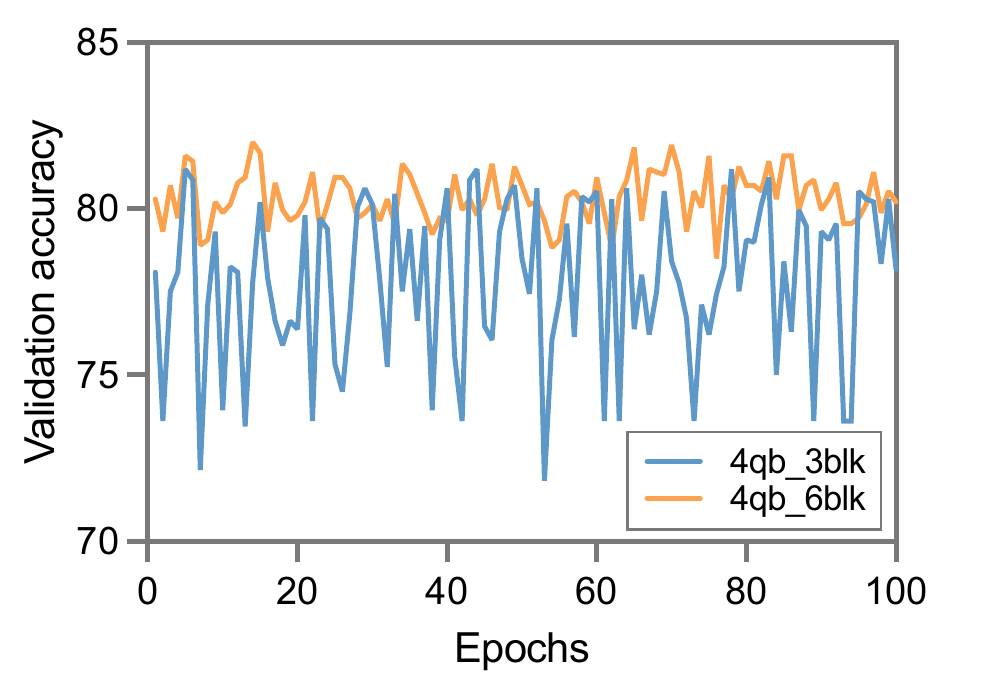}
    }
    \subfigure[Accuracy of 9-qubit QNNs classifiers]{
    \label{fig:small_comp_9qb}
    \includegraphics[width=.6\columnwidth]{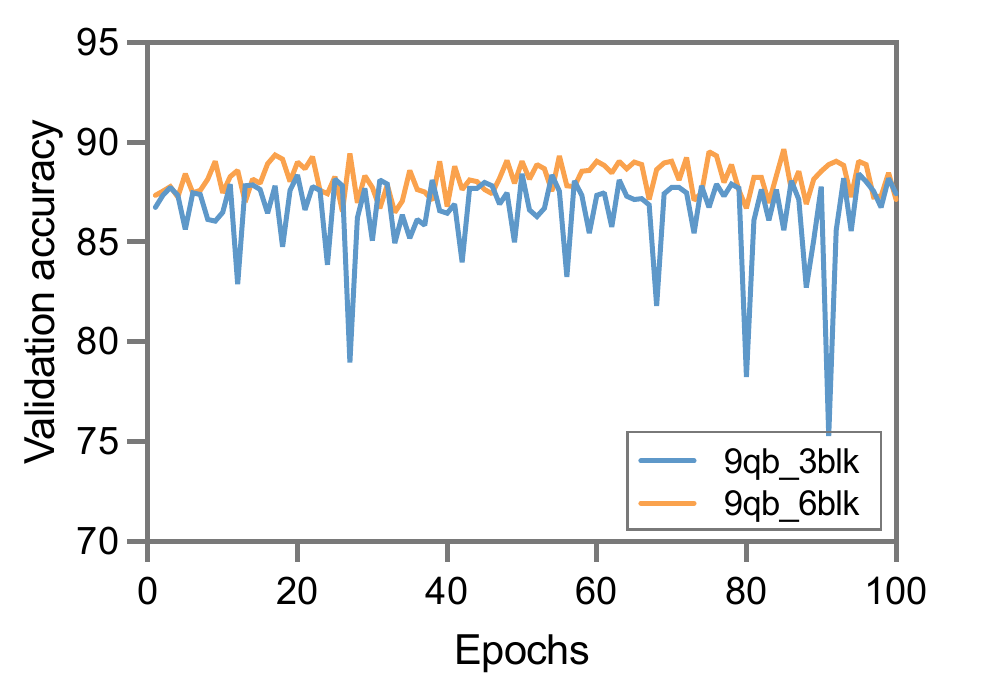}
    } 
    \subfigure[Accuracy of 16-qubit QNNs classifiers]{
    \label{fig:small_comp_16qb}
    \includegraphics[width=.6\columnwidth]{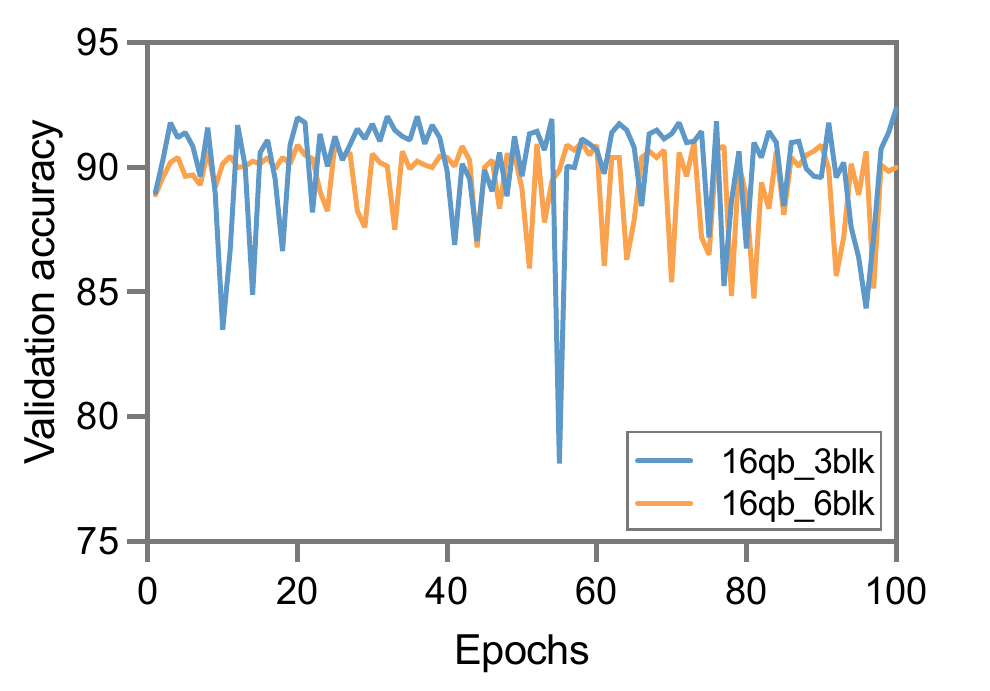}
    }
    \caption{\textbf{The comparison of accuracy between 3-block and 6-block QNN classifiers.} The QNN models with the same amount of data qubits (qb) but the different numbers of variational circuit blocks (blk) are compared. The images demonstrate the models with the same number of qubits achieve similar performance, but the accuracy of the models with six blocks shows less fluctuation.}
    \label{fig:small_qubits_comp}
\end{figure*}

\begin{figure*}[h]
    \centering
    \includegraphics[width=.9\linewidth]{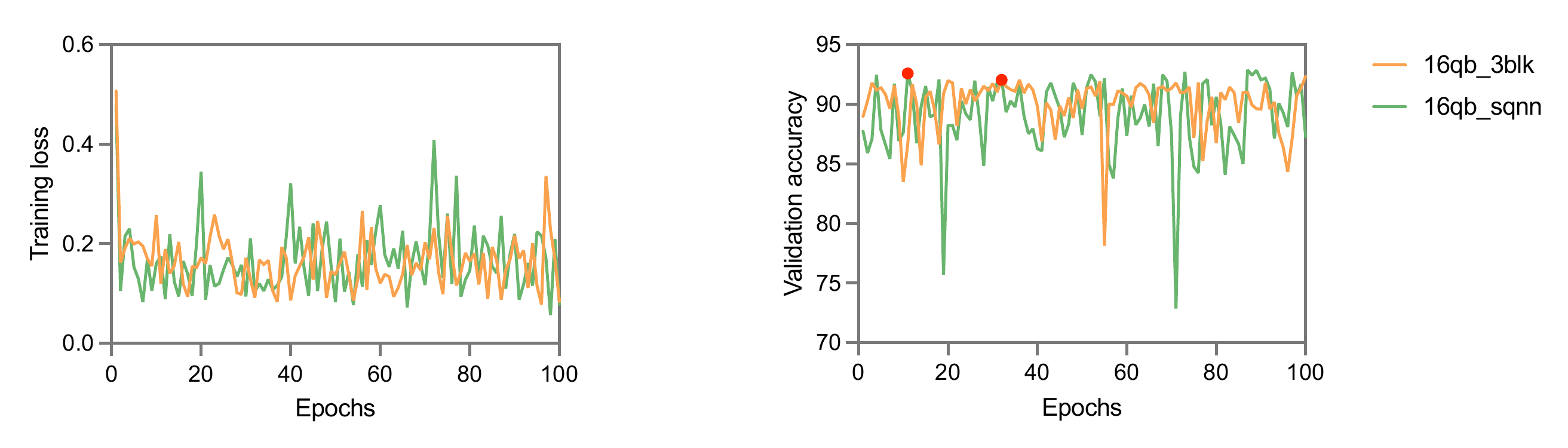}
    \caption{\textbf{The comparison of performance between QNN and SQNN classifiers}. The SQNN and the single-device QNN models both have three variational circuit blocks (blk) and 16 data qubits (qb). Their comparable performance illustrate the effectiveness of the SQNN model.}
    \label{fig:comp_small16_large4_all}
\end{figure*}

\begin{figure*}[h]
    \centering
    \includegraphics[width=.9\linewidth]{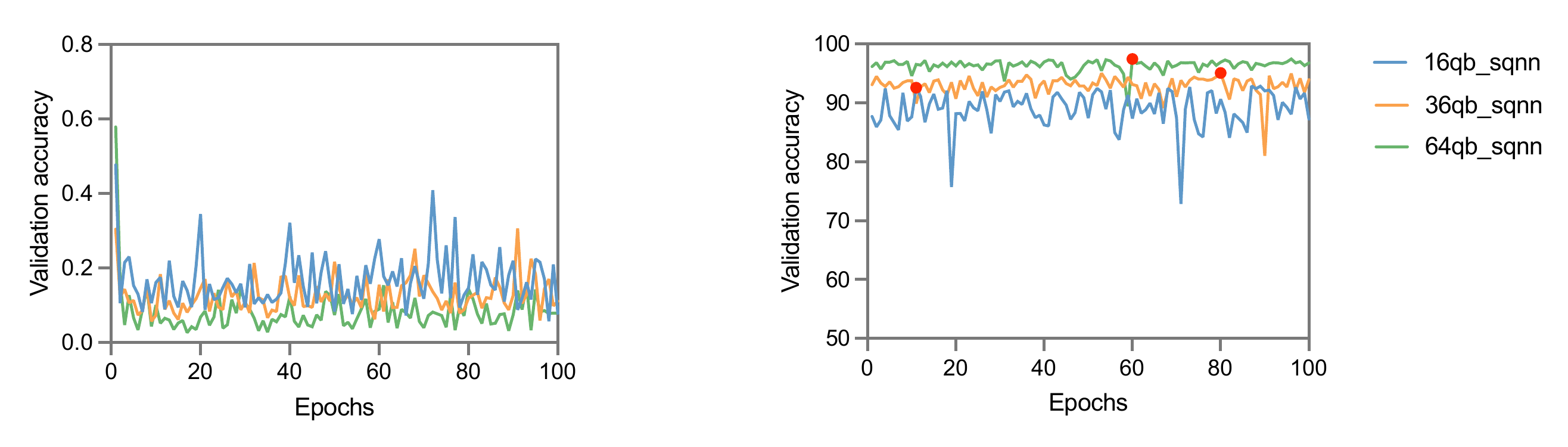}
    \caption{\textbf{The performance of SQNN classifiers in three scales of quantum systems.} Three SQNN models are implemented with 16, 36, and 64 data qubits (qb), and their best accuracy is marked with red dots. The comparison shows that models with more quantum resources can perform better.}
    \label{fig:comp_large_all}
\end{figure*}

We first show some observations about how the hardware limitations of NISQs impact the performance of regular QNN classifiers. Assuming there are three quantum machines of size $4+1$, $9+1$, and $16+1$ qubits, we implement a QNN classifier with the basic model defined above on each device. The last qubit works as the readout, and the remaining qubits are used to load classical training data with $R_x$ gate in encoding unit described in Sec.~\ref{sec:encode}. During the encoding phase, the classical handwritten digit images of size 28 $\times$ 28 are downscaled to $2\times 2$, $3\times 3$, and $4\times 4$ sizes and loaded on the three quantum devices, respectively.

Fig.~\ref{fig:small_3} demonstrates the performance of the QNN classifiers with three blocks. Each model has been trained 100 epochs, and its best accuracy is indicated by a red dot in the right panel of Fig.~\ref{fig:small_3}. Among the QNN classifiers with three blocks, \textit{16qb\_3blk} model with 16 data qubits has the best accuracy of 92.04\% on the validation dataset. The following is the \textit{9qb\_3blk} model with 9 data qubits has an accuracy of 88.37\%. The \textit{4qb\_3blk} model with 4 data qubits has the lowest accuracy of 81.18\%. We also evaluate the performance of QNN classifiers with six blocks on these quantum devices and present the results in Fig.~\ref{fig:small_6}. The best accuracy of the models are 90.99\% on \textit{16qb\_6blk} model, 89.45\% on \textit{9qb\_6blk} model and 82.00\% on \textit{4qb\_6blk} model. We observe that the QNN model that learns from higher dimensional input data can achieve lower training loss and higher accuracy with the same number of blocks based on the comparison among the models with various amounts of data qubits. Additionally, because more parameters need to be trained, the larger model requires more training epochs to reach the point with the best performance.

We further compared the classification accuracy of three pairs of QNN classifiers with the same number of data qubits but the different number of blocks. For the QNNs with the same number of data qubits, as shown in Fig.~\ref{fig:small_qubits_comp}, more blocks will not improve the validation accuracy but can improve the stability of the training process. Therefore, we argue that the number of qubits (the dimension of input data) has a more significant impact on the classification accuracy of the QNN classifier than the circuit depth.

\subsection{SQNN with even partitioning}

This subsection shows the performance of SQNN-based quantum classifiers. SQNN classifiers are supposed to learn from raw-sized training data. However, very high-dimensional quantum operations cannot be simulated by classical devices. As a result, in our simulator, a VQC made up of basic quantum blocks can only contain at most 17 qubits.\\

First, we demonstrate the effectiveness of the SQNN classifier by comparing the performance of the SQNN classifier trained on data of the same size with the regular QNN classifier. We implement an SQNN classifier trained on MNIST ``3" and ``6" images of downscaled size 4 x 4 to compare with the \textit{16qb\_3blk} model that achieves the best performance among regular QNN classifiers. 

Assuming there are five $(4+1)$-qubit quantum devices in a quantum system, we use four of them as the quantum feature extractors and the remaining one as the quantum predictor. The quantum circuits on quantum feature extractors are structured similarly to the \textit{16qb\_3blk} model, and the quantum circuit on the quantum predictor has an encoding unit and a block to incorporate the extracted features. We refer the SQNN model as the \textit{16qb\_sqnn} model, which has 16 data qubits in total. We evenly partition the 4x4 downscaled image into four 2x2 segments, as shown in the left panel of Fig.~\ref{fig:mnist6}. The comparison of the training loss and validation accuracy between the \textit{16qb\_3blk} and \textit{16qb\_sqnn} model is shown in Fig.~\ref{fig:comp_small16_large4_all}. \textit{16qb\_sqnn} achieves the accuracy of 92.59\%, which is comparable with the accuracy achieved by  \textit{16qb\_3blk}, 92.04\%. Thus, the experiment shows the effectiveness of SQNN.\\

Next, we discuss the performance of SQNN classifiers in quantum systems at different scales. SQNN classifiers are implemented in three scales of quantum systems, and their performance is shown in Fig.~\ref{fig:comp_large_all}. Each quantum system considered in this experiment contains a $(4+1)$-qubit quantum predictor and four quantum feature extractors of $(4+1)$-qubit, $(9+1)$-qubit, or $(16+1)$-qubit size, i.e., the data samples are evenly partitioned into four segments as the first panel shown in Fig.~\ref{fig:mnist6}, then the three SQNN classifiers are trained on $4 \times 4$, $6 \times 6$, and $8 \times 8$ sizes of images, respectively. The SQNN classifier with 64 data qubits (four quantum feature extractors with 16 qubits each) in total is denoted as \textit{64qb\_sqnn} model and achieves the best accuracy of 97.47\%. The \textit{36qb\_sqnn} model with 36 data qubits has accuracy of 95.10\% and \textit{16qb\_sqnn} model with 16 data qubits has accuracy of 92.59\%.

Based on the results, one can observe that the SQNN classifiers implemented in the larger-scale quantum systems have better performance, which is consistent with the observations in Sec.~\ref{sec:obs}. The \textit{36qb\_sqnn} model and \textit{64qb\_sqnn} that utilize more quantum resources from multiple quantum devices all achieve higher classification accuracy than \textit{16qb\_3blk} model, the best model trained on a single quantum device.

\subsection{SQNN with uneven partitioning}

\begin{figure*}[t]
    \centering
    \includegraphics[width=.9\linewidth]{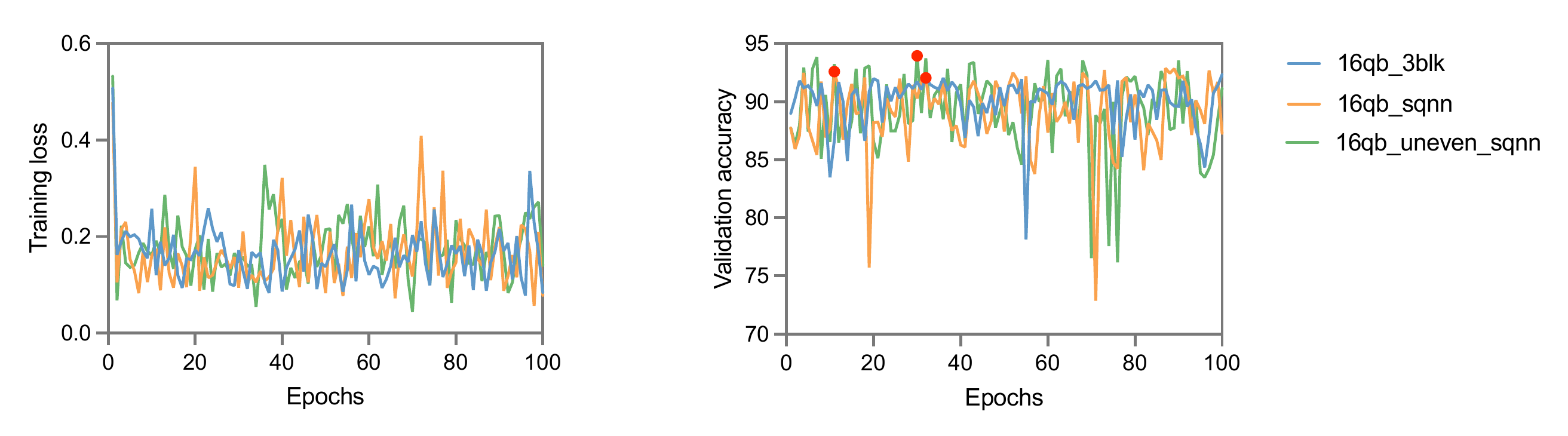}
    \caption{\textbf{The comparison of performance among input partition strategies.} Three 16 data-qubit (qb) models are compared in this experiment: single-device model $16qb\_3blk$, SQNN model $16qb\_sqnn$ and SQNN model with uneven input partition $16qb\_uneven\_sqnn$. The availability of various input partition strategies is proven by their comparable performance. }
    \label{fig:comp_uneven}
\end{figure*}

We now assume a quantum system containing various size quantum feature extractors. A quantum system, for example, contains four quantum devices: a $(8+1)$-qubit quantum device, two $(4+1)$-qubit quantum devices, and a $(3+1)$-qubit quantum device. Typically, the smallest device is chosen as the quantum predictor. To make full use of the quantum resources on quantum feature extractors, we downscale the data to $4 \times 4$ and partition it into a $2 \times 4$ segment and two $2 \times 2$ segments, as the second panel in Fig.~\ref{fig:mnist6}. We then respectively assign them to the $(8+1)$-qubit quantum device and the two $(4+1)$-qubit quantum devices. Fig.~\ref{fig:comp_uneven} compares the performance of uneven partitioning model \textit{16qb\_uneven\_sqnn} to the other two models, \textit{16qb\_3blk} and \textit{16qb\_sqnn}, that are trained with the same size input data. Fig.~\ref{fig:comp_uneven} illustrates that the SQNN with an uneven partitioning strategy performs similarly to the other two models. Hence, we argue that SQNN in the quantum system with quantum devices of different sizes is also effective.

\section{Related Work}\label{sec:related}

The topic studied in this paper is known as quantum neural network \cite{altaisky2001quantum, jeswal2019recent, ricks2003training}, a case of variational quantum algorithm \cite{cerezo2021variational}. There have been some works in this field that perform classification tasks with quantum computing\cite{abohashima2020classification}. The approaches include quantum machine learning \cite{willsch2020support, rebentrost2014quantum, schuld2016prediction, da2016quantum}, quantum-inspired machine learning\cite{tiwari2019towards, sergioli2019new, ding2021quantum}, and hybrid quantum-classical machine learning \cite{adhikary2020supervised, schuld2019quantum, chalumuri2021hybrid, gianelle2022quantum, tao2022laws}.

Among these approaches, the hybrid approach attracts much attention. A quantum multi-class classifier, QMCC, proposed in \cite{chalumuri2021hybrid}, uses $n$ qubits to encode an N-dimensional data vector and uses $k$ ancilla qubits as readouts to store the predicted label out of $k$ classes. Although the quantum computer is much faster than the classical computer, the speed of hybrid quantum-classical approaches is still constrained by classical phases. Hence, some efforts have been made for acceleration. A ``single-shot training" style proposed in \cite{adhikary2020supervised} uses all the input samples with the same label to train the classifier at the same time to speed up the training procedure. QReliefF accelerates the training process by combining quantum machine learning and edge computing so that some computations can be finished parallelly \cite{liu2020quantum}. Besides the quantum computation phase, the time cost in the quantum encoding phase can also be reduced by using a quantum dataset so that the quantum devices do not need to perform encoding for classical data. A quantum dataset NTangled is proposed in \cite{schatzki2021entangled}. It is composed of quantum states with different amounts and types of entanglements.

Quantum models are limited in their ability to investigate potential patterns in data due to hardware restrictions on quantum devices. The existing approaches must shrink the raw data size to fit the number of available qubits on the quantum device. The intuitive method is to downscale the raw data into a small size, but much helpful information is lost. A solution proposed in \cite{chen2020hybrid} is to use a quantum-inspired tensor network (TN), particularly a matrix product state, as a feature extractor to reduce data size and then use the compressed data as the input of VQC to perform supervised learning tasks. However, the quantum-inspired TN runs on classical devices, which slows down the training speed. A large-scale QML is proposed in \cite{haug2021large}. It computes quantum kernels using randomized measurements and loads high-dimensional classical data on a quantum device by repeating the encoding layer, which increases the depth of the quantum circuit and is incompatible with the QNN approaches considered in our work.

The distributed systems of QNN are designed to train a quantum model among multiple parties collaboratively. QuantumFed \cite{xia2021quantumfed} combines QNN and federated learning to allow multiple quantum devices to train a QML model collaboratively on their private dataset. And the authors further consider the security issue of quantum federated learning. Their work presented in \cite{xia2021defending} protects the QML model trained in the federated learning system from Byzantine attacks, in which adversaries upload malicious information to degrade model performance. The federated QNN through blind quantum computing implemented with differential privacy in order to guarantee clients' privacy \cite{li2021quantum}. DSQML, distributed secure quantum machine learning, is proposed to implement a secure QNN without leakage of any information about training data \cite{sheng2017distributed}. The above approaches are computationally intensive on classical devices and do not overcome the quantum hardware limitation. In comparison to the previous works, our proposed approach performs no additional computation on classical devices other than the update calculation required in the hybrid method.

\section{Conclusion}\label{sec:concl}

Considering the difficulty and cost of building a large-scale quantum device for QNN, we propose a scalable quantum neural network (SQNN), which is deployed on multiple small-size quantum devices. In this paper, we consider SQNN models for a binary classification task. The SQNN classifier learns from high-dimensional training data by partitioning it into small segments. And the resource-constrained quantum devices in the quantum system extract local features from the segments in parallel. The extracted features are then merged and predicted using a quantum device acting as a quantum predictor. Moreover, the training sample could be partitioned into segments of different sizes to fully utilize the quantum devices with various amounts of quantum resources. Furthermore, we show that a single small-size quantum device is also able to train a large neural network according to the SQNN approach. Extensive experiments indicate that having more quantum resources improves SQNN classifier accuracy significantly, and our proposed approach outperforms regular QNNs trained with limited quantum resources.

\section*{Acknowledgment}
The authors would like to thank all the reviewers for their helpful comments. 
This project was supported in part by US National Science Foundation grant CNS-1816399. This work was also supported in part by the Commonwealth Cyber Initiative, an investment in the advancement of cyber R\&D, innovation and workforce development. For more information about CCI, visit cyberinitiative.org.

\bibliographystyle{IEEEtran}
\bibliography{ref}

\end{document}